\documentclass[letterpaper]{article} 
\usepackage{aaai2027}
\usepackage[hyphens]{url}  
\usepackage{graphicx} 
\usepackage{natbib}  
\usepackage{caption} 
\usepackage{algorithm}
\usepackage{algorithmic}
\usepackage{times}  
\usepackage{helvet}  
\usepackage{courier}  
\usepackage[hyphens]{url}  
\usepackage{graphicx} 
\usepackage{threeparttable}
\usepackage{comment}
\usepackage{booktabs}
\usepackage{multirow}
\usepackage{natbib}  
\usepackage{caption} 
\usepackage{tabularray}
\usepackage{booktabs}    
\usepackage{multirow}    
\usepackage{array}       
\usepackage{threeparttable} 
\usepackage{algorithm}
\usepackage{algorithmic}
\usepackage{amsmath, amssymb}
\usepackage{xcolor}
\usepackage{color}
\usepackage[table]{xcolor}
\usepackage{newfloat}
\usepackage{listings}
\DeclareCaptionStyle{ruled}{labelfont=normalfont,labelsep=colon,strut=off} 
\floatstyle{ruled}
\newfloat{listing}{tb}{lst}{}
\floatname{listing}{Listing}

\usepackage{booktabs}

\newcommand{\corresponding}{\textsuperscript{\textdagger}}
\title{Aligning the Incomplete: Joint Distribution Calibration for Multimodal EEG-Eye Emotion Recognition}

\author{
    Yang Wu\textsuperscript{\rm 1},
    Jinpeng Li\textsuperscript{\rm 1}\corresponding
}
\affiliations{
    \textsuperscript{\rm 1} School of Automation Science and Engineering,\\
    South China University of Technology
}

\begin{document}

\maketitle

\begin{abstract}
The success of cross-subject multimodal emotion recognition hinges on maintaining the consistency of the joint data distribution across individuals. However, real-world deployment frequently triggers the \emph{asymmetric joint distribution collapse}: EEG signals suffer from severe cross-subject distribution shifts, while eye movements sensors are susceptible to packet loss and tracking failures. Existing methods treat domain adaptation and missing-modality imputation as disjoint tasks. Consequently, they fail to resolve the compounded errors when both degradations co-occur, either propagating domain shifts through imputed signals or destroying the joint decision boundary. To tackle this unified challenge, we propose GUARD (\textbf{G}radient-guided \textbf{U}nsupervised \textbf{A}symmetric \textbf{R}ecovery of \textbf{D}istributions). First, GUARD establishes a reliable anchor manifold in the source domain by employing a theoretically grounded gradient-weighted objective, which forces the robust EEG modality to preemptively entangle task-discriminative ocular features. Next, to structurally recover the collapsed joint distribution, we constrain a generative module with downstream perceptual losses, prioritizing emotion-discriminative semantics over mere signal fidelity. Finally, we formulate target-domain adaptation as an ill-posed inverse problem. By driving a cycle-consistent flow, we achieve unsupervised calibration of the recovered joint distribution directly on the target-domain manifold. Extensive experiments demonstrate that GUARD significantly outperforms state-of-the-art methods, maintaining resilient discriminative performance even under complete auxiliary modality failure. Our code and models are made publicly available to ensure complete reproducibility.

\end{abstract}


\section{Introduction}

Emotion recognition promises naturalistic human-computer interaction and holds substantial societal value for mental health monitoring~\cite{affective2024survey, anxietyreview2025}. Among various physiological signals, electroencephalogram (EEG) and eye movements  have emerged as pivotal modalities due to their complementary characteristics: EEG signals objectively reflect internal neural activity, while eye movements provide fine-grained behavioral cues~\cite{zheng2014}. Their bimodal fusion has consistently achieved state-of-the-art performance on standardized benchmarks such as the SEED series datasets~\cite{Lu2015, Jiang2025}.

\begin{figure}[t]
\centering
\includegraphics[width=1\linewidth]
{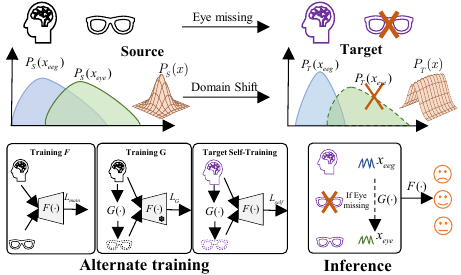}  
 \caption{\textbf{Overview of the GUARD framework.} (Top) Asymmetric joint distribution collapse caused by concurrent domain shift and missing modality. (Bottom) Alternate training paradigm to recover and calibrate the joint distribution under complete auxiliary modality failure.}
\label{conclusion}
\end{figure}
Despite these advances, multimodal affective computing systems encounter a bottleneck: \emph{the cross-subject domain shift, which is severely exacerbated when eye movements are lost during inference due to ambient illumination and environmental factors.} These two bottlenecks have been treated as independent problems: domain adaptation aims to minimize the distribution discrepancy between labeled source subjects and unlabeled target individuals~\cite{MMD, CFDA-CSF, CSMM}, while missing-modality imputation leverages generative models to synthesize the absent signal stream~\cite{MMIN, VAES, DFM, UMAP}. However, in practice, these two challenges are not independent---they converge to trigger a strong degradation effect that is mutually reinforcing rather than merely co-occurring.

We propose to view this coupled problem through \emph{joint distribution distortion}, as conceptually illustrated in the top half of Fig.\ref{conclusion}. In the source, we have access to the complete joint distribution \(P_\mathcal{S}(x_{eeg}, x_{eye})\). Under standard domain adaptation assumptions, domain shift is characterized by the marginal discrepancy \(P_\mathcal{S}(X_{eeg}, x_{eye}) \neq P_\mathcal{T}(x_{eeg}, x_{eye})\) when full modalities are available. However, when the eye movements modality is temporarily absent in the target domain, the distribution collapses to the incomplete marginal \(P_\mathcal{T}(x_{eeg})\)—the joint distribution is degraded, losing both discriminative information and the structural dependencies between modalities. This \emph{joint distribution collapse} renders conventional domain-agnostic completion strategies inadequate: they inadvertently synthesize signals that increase semantic confusion with source-domain knowledge rather than facilitating emotion discrimination in the target domain. Critically, the two challenges are coupled: \emph{modality missing not only removes discriminative information but also amplifies the effective domain shift by distorting the joint distribution manifold that the classifier relies on.}

To address this unified problem, the proposed GUARD involves three interconnected components. First, we introduce the Gradient-MSE (G-MSE) that computes sample-adaptive channel importance from the classifier's gradient. Grounded in first-order Taylor expansion, this loss adaptively assigns higher reconstruction weights to channels with larger gradient magnitudes in a sample-dependent manner, steering representational capacity toward task-discriminative dimensions and establishing a robust anchor manifold.  Second, to recover the complete joint distribution when eye movements  is missing, we train a generator that synthesizes semantically meaningful ocular features that maximize emotion discriminability rather than merely minimizing signal fidelity. Third, to calibrate the recovered joint distribution against the domain shift across subjects, we formulate semantic completion as an implicit \emph{equation-solving problem within a self-supervised flow}. Given the target EEG as a known anchor, we impose cycle-consistency on the cross-modal reconstruction to recover the implicit eye features. This achieves unsupervised target-domain feature calibration, ensuring the generated representations are semantically cohesive with target brain oscillations and well-aligned with the classifier's decision boundary. Our contributions include:
\begin{enumerate}
    \item We propose GUARD, a unified framework that conceptualizes cross-subject adaptation and auxiliary modality missing not as disjoint tasks, but as a coupled \emph{asymmetric joint distribution collapse.} We tackle the resulting cascade error through a novel joint distribution recovery and alignment paradigm.
    \item We introduce a theoretically grounded G-MSE loss that dynamically assigns sample-adaptive channel importance. By unifying classification and reconstruction objectives, it forces the primary modality to preemptively entangle task-discriminative cross-modal semantics.
    \item We design a cycle-consistent self-supervised flow that formulates target-domain semantic completion as an ill-posed equation-solving problem, enabling fully unsupervised feature calibration to recover the optimal joint distribution without relying on target ocular labels.
    \item We achieve new state-of-the-art results on the SEED series benchmarks, providing a robust baseline that maintains highly resilient physiological emotion decoding even under complete auxiliary modality failure.
\end{enumerate}

\section{Related Work}

\textbf{Cross-Subject Domain Adaptation.} Cross-subject domain adaptation (DA) aims to mitigate distribution shifts across subjects by aligning source and target feature representations. Early attempts focused on minimizing statistical divergence using metrics such as MMD \cite{MMD} and CORAL \cite{CORAL}. Subsequently, adversarial learning-based frameworks have dominated the field, with recent advances introducing multi-task adversarial adaptation \cite{MTADA2025}, multi-source adaptive gating \cite{MSAGDA2025}, dynamic distribution alignment \cite{UDADDA2025}, and semi-supervised domain adversarial learning \cite{SEDA2025} to enhance cross-subject generalization. Domain generalization strategies have also been explored to learn subject-invariant features without accessing target data \cite{MMASEDG2025, LATN2026}. Meanwhile, multimodal emotion recognition has been shown to significantly outperform unimodal approaches \cite{MMDA, CFDA-CSF, Jiang2025}. \textit{However, these DA paradigms inherently assume symmetric modality structures, requiring source and target domains to share identical complete modalities.} This assumption renders them fragile when target subjects face unpredictable modality deficiencies.

\textbf{Missing Modality Imputation.} Traditional approaches treat imputation as a matrix completion problem using low-rank factorization. With deep learning, generative models have been widely adopted to synthesize missing streams. Diffusion-based approaches have recently emerged as a powerful paradigm \cite{FedDISC2025}, while progressive learning frameworks \cite{ProLF2025} and generative random modality dropout \cite{RMDG2025} have enhanced robustness under various absence scenarios. Cross-modal graph attention networks \cite{MICGA2025} have also been proposed to handle incomplete modalities. \textit{However, most imputation methods assume training and testing data are identically distributed.} They degrade significantly when applied to new subjects with distinct physiological traits.

Despite advances in imputation and DA, no framework jointly tackles structural recovery and distribution shift. Existing methods adopt disjoint pipelines, which propagate imputation errors and yield suboptimal solutions \cite{che2018recurrent}. GUARD formulates missing modality completion and cross-subject adaptation as a unified unsupervised calibration problem over an asymmetric joint distribution, achieving robust generalization even under severe modality deficiency.

\begin{figure*}[h]
\centering
\includegraphics[width=1\linewidth]{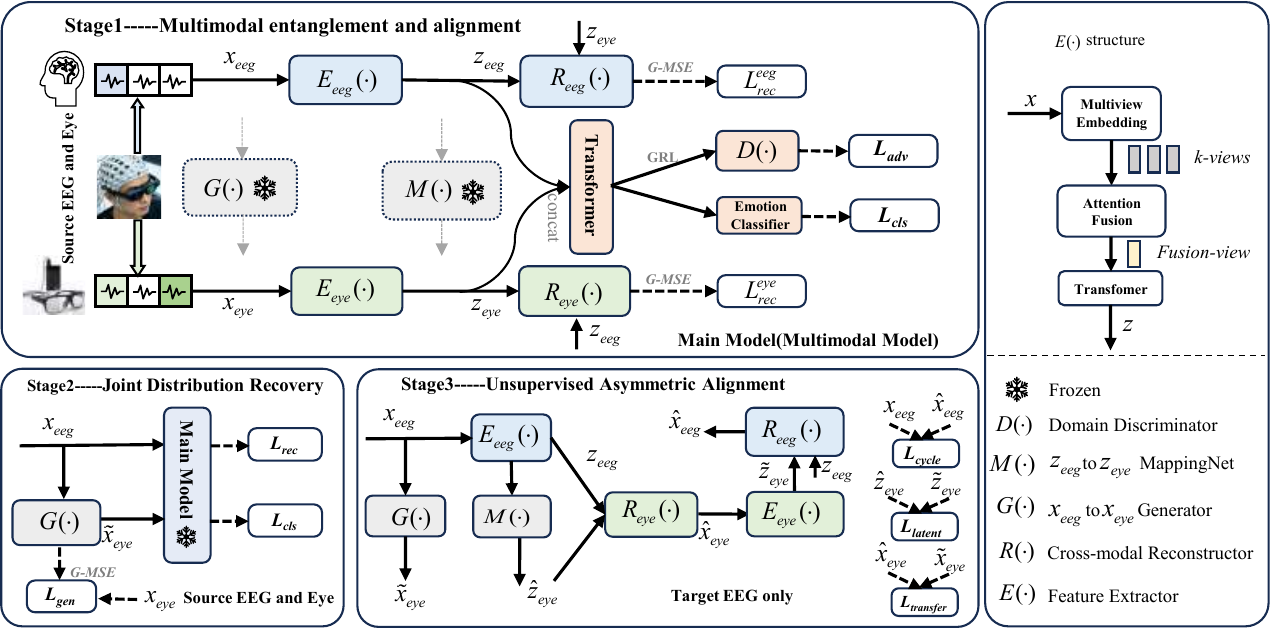} 
\caption{\textbf{Overall architecture and training/inference pipeline of GUARD.} Training proceeds via three alternating stages: (1) multimodal classification model training; (2) generator training with the main model frozen; (3) unsupervised fine-tuning of the full framework using target EEG only. }
\label{GUARD
}
\end{figure*}

\section{Method}
\subsection{A. Problem Formulation}
Traditional cross-subject emotion recognition often treats domain shift and missing modalities as orthogonal challenges, aiming to symmetrically align the marginal distributions $P_{\mathcal{S}}(x) \neq P_{\mathcal{T}}(x)$. In real-world scenarios, however, these issues are profoundly coupled, culminating in a severe \textit{asymmetric joint distribution collapse}.

Let $x_{eeg} \in \mathcal{X}_{eeg}$ and $x_{eye} \in \mathcal{X}_{eye}$ denote the primary (EEG) and auxiliary (eye movements ) signals, respectively. The optimal multimodal classifier relies on the joint distribution $P(x_{eeg},x_{eye})$ factorized as:
\begin{equation}
P(x_{eeg}, x_{eye}) = P(x_{eeg}) \cdot P(x_{eye} \mid x_{eeg})
\end{equation}
During practical target-domain inference, this joint distribution suffers from two concurrent and mutually reinforcing degradations:

\noindent\textbf{1. Structural Degradation (Modality Missing).} The unpredictable loss of eye movements sensors collapses the target joint distribution into a deficient marginal distribution, i.e., $P_{\mathcal{T}}(x_{eeg}, x_{eye}) \rightarrow P_{\mathcal{T}}(x_{eeg})$. The critical structural dependency, represented by the conditional distribution $P_{\mathcal{T}}(x_{eye} \mid x_{eeg})$, is entirely obliterated.

\noindent\textbf{2. Distributional Degradation (Domain Shift).} Individual variability induces an domain shift, afflicting both the marginal distributions ($P_{\mathcal{S}}(x_{eeg}) \neq P_{\mathcal{T}}(x_{eeg})$) and the cross-modal mappings ($P_{\mathcal{S}}(x_{eye} \mid x_{eeg}) \neq P_{\mathcal{T}}(x_{eye} \mid x_{eeg})$). Direct imputation using source-domain knowledge will inevitably propagate and amplify this shift.

This leads to an \emph{asymmetric alignment task}: learning a robust classifier by aligning the complete source joint distribution $P_{\mathcal{S}}(x_{eeg}, x_{eye})$ with the structurally collapsed and distributionally shifted target marginal $P_{\mathcal{T}}(x_{eeg})$. 

To achieve this, GUARD takes three steps:

\noindent\textbf{1. Multimodal entanglement to optimize the anchor.} To prevent the target marginal $P_{\mathcal{T}}(x_{eeg})$ from being semantically vacuous, we must establish a robust baseline. During source training, we force the EEG marginal representations to entangle task-discriminative cross-modal features, creating a resilient anchor manifold for $P_{\mathcal{S}}(x_{eeg})$.

\noindent\textbf{2. Latent joint distribution recovery to model the conditional.} To physically reconstruct the collapsed joint structure, we must explicitly model the lost conditional term $P(x_{eye} \mid x_{eeg})$. We approximate this via a generative mapping, constrained not by mere signal fidelity, but by task-perceptual losses to maximize emotion discriminability.

\noindent\textbf{3. Unsupervised asymmetric alignment to calibrate the shift.} Because $P_{\mathcal{S}}(x_{eye} \mid x_{eeg}) \neq P_{\mathcal{T}}(x_{eye} \mid x_{eeg})$, the generated target joint distribution remains misaligned. We formulate this discrepancy as an \emph{overdetermined equation-solving problem}, driving a cycle-consistent self-supervised flow to unsupervisely calibrate the recovered target joint space against the source anchor manifold.

\subsection{B. Multimodal entanglement and alignment}
Since the target marginal \(P_{\mathcal{T}}(x_{eeg})\) is inherently prone to semantic collapse without its ocular counterpart, we prospectively condition the primary EEG encoder to serve as an information-anchoring hub during source training. To this end, we design a hierarchical Transformer-based classification backbone integrated with a cross-modal reconstructor. Multi-view \cite{Jiang2025}embedding modules first project each raw input modality \(x_{eeg}\) and \(x_{eye}\) into \(K\) latent vectors \(\{H^{(1)},\dots,H^{(K)}\}\), which are then aggregated via an attention-based fusion mechanism to produce a compact discriminative descriptor \(J\):

\begin{equation}
\alpha_k = \frac{\exp((H^{(k)})^\top q)}{\sum_{j=1}^{K}\exp((H^{(j)})^\top q)},\quad J = \sum_{k=1}^{K}\alpha_k H^{(k)}
\end{equation}

The resulting modality-specific representations \(J_{eeg}\) and \(J_{eye}\) are then processed by dedicated transformer encoders to yield enhanced features \(z_{eeg}\) and \(z_{eye}\). These are concatenated and fed into a transformer encoder\cite{vaswani2017attention} to further model the sequential cross-modal dependencies, whose output is finally passed to an MLP classifier \(C(\cdot)\) optimized under the cross-entropy loss \(\mathcal{L}_{cls}\).

Central to our design is a cross-modal reconstructor\(\mathcal{R}(\cdot)\), with dual roles: capturing complementary information during source training, and establishing a reliable anchor for self-supervised calibration. The reconstructor is implemented as a Transformer decoder that takes the source modality representation as queries and the target modality representation as keys/values, producing the reconstructed signal via cross-attention and feedforward neural networks.  To this end, we introduce a gradient-weighted reconstruction strategy that assigns sample-adaptive, channel-wise weights based on each channel's influence on the model's decision boundary. We formalize this by abstracting the multimodal network as a nonlinear function \(F:\mathbb{R}^d \to \mathbb{R}^C\), with \(x = (x^1,\dots,x^d) \in \mathbb{R}^d\) denoting the multi-channel input. For a given sample \(x_0\), the first-order Taylor expansion of \(F\) about \(x_0\) yields:

\begin{equation}
F(x) = F(x_0) + \nabla F(x_0)^\top (x - x_0) + \mathcal{O}(\|x - x_0\|^2)
\end{equation}

As the model converges, the higher-order residuals vanish under the convergence condition \(\|x - \tilde{x}\|^2 \to 0\), yielding the linear approximation:

\begin{equation}
F(x) - F(\tilde{x}) \approx \nabla F(x_0)^\top (x - \tilde{x}) = \sum_{c=1}^{d} \frac{\partial F}{\partial x^c}\bigg|_{x_0} (x^c - \tilde{x}^c)
\end{equation}

A complete proof of this asymptotic convergence is provided in the supplementary material.

This decomposition reveals a critical insight: the influence of reconstruction error on the final prediction is linearly modulated by the gradient magnitude \(|\partial F / \partial x^c|\). Accordingly, we define \( w_{i,c} \) as the importance weight for the \( c \)-th channel of the \( i \)-th sample:

\begin{equation}
w_{i,c} = \left|\frac{\partial \mathcal{L}}{\partial x_{i,c}}\right|,\quad \mathcal{L} = \mathcal{L}_{cls} + \mathcal{L}_{MSE}
\end{equation}
where \(\mathcal{L}_{MSE}\) denotes the standard unweighted reconstruction loss, which ensures basic signal fidelity.  

To counteract the volatility of gradient estimates, we maintain an Exponential Moving Average (EMA) buffer with momentum \(m = 0.9\). The resultant Gradient-Weighted Mean Squared Error (G-MSE) \(\mathcal{L}_{G-MSE}\) is formulated as:

\begin{equation}
\mathcal{L}_{G-MSE} = \frac{1}{N}\sum_{i=1}^{N}\sum_{c=1}^{d} w_{i,c} \cdot (x_{i,c} - \tilde{x}_{i,c})^2
\end{equation}

Concurrently, to suppress subject-specific biases and address the marginal shift, we introduced domain adversarial training through the gradient reversal layer.
\begin{equation}
\mathcal{L}_{adv} = -\frac{1}{N}\sum_{i=1}^{N}\sum_{k=1}^{N_{domain}} d_{i,k} \log(\hat{d}_{i,k})
\end{equation}

The complete source-domain objective thus consolidates classification, entanglement, and adversarial alignment:

\begin{equation}
\mathcal{L}_{main} = \mathcal{L}_{cls} + \alpha \mathcal{L}_{G-MSE} + \beta\mathcal{L}_{adv}
\end{equation}

Through this joint optimization, the source domain yields a strongly entangled and subject-invariant joint manifold—establishing the robust anchor that underpins all subsequent recovery operations.

\subsection{C. Latent Joint Distribution Recovery}

With the source anchor established, structural recovery of the target joint distribution reduces to estimating the lost conditional dependency \(P(x_{eye} \mid x_{eeg})\). As noted in Section A, the ideal joint distribution factorizes accordingly. Since only the target marginal \(P_{\mathcal{T}}(x_{eeg})\) is observable, recovery hinges entirely on reliable characterization of this conditional relationship. To this end, we introduce a cross-modal MLP generator \(G(\cdot)\) that explicitly learns the mapping from EEG to the eye movement space.

Keeping the pre-trained main model completely frozen, we train \(G(\cdot)\) to synthesize pseudo eye movements signals \(\hat{x}_{eye} = G(x_{eeg})\). A physical-space consistency constraint is imposed as:

\begin{equation}
\mathcal{L}_{gen} = \frac{1}{N}\sum_{i=1}^{N}\sum_{c=1}^{d_w} w_{i,c} \cdot (x_{i,c}^{eye} - \hat{x}_{i,c}^{eye})^2
\end{equation}

However, fidelity to the raw signal space alone is insufficient for discriminative tasks. The synthesized signals are therefore propagated through the frozen main model to impose downstream perceptual supervision. This yields the generator objective:

\begin{equation}
\mathcal{L}_G = \mathcal{L}_{gen} + \eta(\mathcal{L}_{cls} +  \mathcal{L}_{G-MSE})
\end{equation}

Both \(\mathcal{L}_{cls}\) and \(\mathcal{L}_{G-MSE}\) are evaluated on synthesized ocular signals via the frozen classifier and reconstructor, compelling the generator to prioritize discriminative semantics over signal fidelity. Crucially, freezing all other modules anchors the recovered conditional distribution to the source manifold, preventing spurious drift—a design validated in the supplementary material.

\subsection{D. Unsupervised Asymmetric Alignment}

While the generator successfully recovers the form of the conditional distribution, it is intrinsically biased toward the source condition \(P_\mathcal{S}(X_{eye} \mid X_{eeg})\). The target domain, however, exhibits a distinct conditional shift \(P_\mathcal{S} \neq P_\mathcal{T}\), yet provides no ground-truth ocular labels for direct adaptation. We address this dilemma by recasting target-domain calibration as an ill-posed inverse problem, leveraging the well-grounded source joint manifold as a natural regularizer. We denote the cross-modal reconstruction pipeline as a deterministic composite mapping \(\mathcal{W}\):

\begin{equation}
\begin{split}
(\tilde{x}_{eeg}, \tilde{x}_{eye}) = \big( & \mathcal{R}_{eeg}(z_{eeg}, z_{eye}), \\
& \mathcal{R}_{eye}(z_{eye},z_{eeg}) \big) \triangleq \mathcal{W}(x_{eeg}, x_{eye})
\end{split}
\end{equation}

In the target domain, only \(x_{eeg}\) is observed, with the reconstruction target naturally set as \(\hat{x}_{eeg} = x_{eeg}\). Substituting this constraint into the composite mapping yields an equation parameterized by the unknown \(\hat{x}_{eye}\):

\begin{equation}
\hat{x}_{eeg} = \mathcal{W}_{eeg}(x_{eeg}, \hat{x}_{eye})
\end{equation}
where \(\mathcal{W}_{eeg}\) denotes the EEG-reconstruction component. This constitutes an overdetermined system that is both highly nonlinear and severely ill-posed, precluding any closed-form solution.

To render this tractable, we introduce a latent MLP mapping network \(\mathcal{M}(\cdot)\) that generates an implicit ocular representation conditioned on the target EEG:

\begin{equation}
\hat{z}_{eye} = \mathcal{M}(z_{eeg}), \quad \hat{x}_{eye} = \mathcal{R}_{eye}(\hat{z}_{eye}, z_{eeg})
\end{equation}

The overdetermined constraint is then relaxed into a cyclic reconstruction objective:

\begin{equation}
\mathcal{L}_{cycle} = \|\mathcal{W}_{eeg}(x_{eeg}, \hat{x}_{eye}) - x_{eeg}\|^2
\end{equation}

Nevertheless, the deep nonlinearity of \(\mathcal{W}\) renders this objective non-convex, and naive optimization risks converging to physically implausible solutions. To confine the search within a physiologically meaningful subspace, we introduce a Lagrangian-style consistency regularizer\cite{Zhou2024Augmented}:

\begin{equation}
\mathcal{L}_{latent} = \|\hat{z}_{eye} - E_{eye}(\hat{x}_{eye})\|^2
\end{equation}

This regularizer enforces a closed-loop consistency between the generated and re-encoded implicit representations, effectively projecting the optimization onto a manifold that preserves the structural dependencies inherent to the true joint distribution. A subsequent knowledge distillation loss transfers the refined representation back to the generator:

As detailed in the supplementary material, this regularizer relaxes the hard constraint \(\hat{z}_{eye} = E_{eye}(\hat{x}_{eye})\) into a least-squares penalty, projecting the optimization onto a physiologically meaningful manifold—ensuring closed-loop consistency and mitigating the ill-posedness of the underdetermined system. A knowledge distillation loss then transfers the refined representation back to the generator:

\begin{equation}
\mathcal{L}_{transfer} = \|G(x_{eeg}) - \hat{x}_{eye}\|^2
\end{equation}

The complete self-supervised calibration objective consolidates these constraints:

\begin{equation}
\mathcal{L}_{self} = \mathcal{L}_{cycle} + \gamma \mathcal{L}_{latent} +  \mathcal{L}_{transfer}
\end{equation}
Through this, the recovered joint distribution is unsupervisedly calibrated on the target manifold, producing ocular representations that are semantically aligned with target EEG and the classifier's decision boundary—without requiring any ground-truth target eye movements data.

At inference, complete target data bypass the generator and are directly processed by the multimodal model, whereas data with missing eye movements are first reconstructed by the generator before being classified by the multimodal model.

\section{Experiments}
\begin{table*}[t]
\centering

\renewcommand{\arraystretch}{1.1}
\setlength{\tabcolsep}{7pt}
\footnotesize

\begin{tabular}{lcccccclcccc}
\toprule

 \multirow{2}{*}{Multimodal} &\multicolumn{2}{c}{SEED} & \multicolumn{2}{c}{SEED-IV} & \multicolumn{2}{c}{SEED-V} &\multirow{2}{*}{Missing-Modal} & \multicolumn{2}{c}{SEED} & \multicolumn{2}{c}{SEED-IV} \\
\cmidrule(lr){2-3} \cmidrule(lr){4-5} \cmidrule(lr){6-7} \cmidrule(lr){9-10} \cmidrule(lr){11-12}
 & Mean & Std & Mean & Std & Mean & Std &  & Mean & Std & Mean & Std \\
\midrule
DCCA      & 85.92 & 9.61  & 60.89 & 10.91 & 51.57 & 14.77 & TCA       & 63.64 & 14.88 & 56.56 & 13.77 \\
JDA       & 86.54 & 8.48  & 79.36 & 9.45  & 79.35 & 16.40 & DGCNN     & 79.95 & 9.02  & 52.82 & 9.23  \\
MWACN     & 87.50 & 11.96 & 73.89 & 10.03 & 81.48 & 16.23 & DANN      & 75.08 & 11.18 & 47.59 & 10.01 \\
CFDA-CSF  & 90.04 & 9.38  & 87.40 & 9.49  & 90.48 & 13.58 & RGNN      & 85.30 & 6.72  & 73.84 & 8.02  \\
MMDA      & 90.87 & 10.32 & 83.32 & 9.51  & 91.27 & 12.16 & BiDANN    & 83.28 & 9.60  & --    & --    \\
CMSLNet   & --    & --    & 83.15 & --    & 87.32 & --    & BiDANN-S  & 84.14 & 6.87  & 65.59 & 10.39 \\
MHSA      & --    & --    & 83.15 & 9.84  & --    & --    & BiHDM     & 85.40 & 7.53  & 69.03 & 8.66  \\
CSMM     & 94.96 & \textbf{5.27}& 89.82 & 6.22  & 89.22 & 9.59  & PGCN      & 84.59 & 8.68  & 73.69 & \textbf{7.16}\\
CMGNN     & --    & --    & 90.21 & \textbf{3.49}& --    & --    & UDDA      & 88.10 & 6.54  & 73.14 & 9.43  \\
MACDB     & 86.68 & 11.07 & 85.03 & 8.78  & 86.48 & 13.18 & BGAGCN-MT & 89.66 & \textbf{4.72}& 75.78 & 8.17  \\
\midrule
\multicolumn{1}{l}{\textbf{GUARD(Ours)}} & \textbf{94.99} & 5.88& \textbf{92.26} & 6.46& \textbf{95.30} & \textbf{7.50} & \textbf{GUARD(Ours)}& \textbf{91.06} & 6.72& \textbf{77.33} & 7.84\\
\bottomrule
\end{tabular}

\caption{We adopt Leave-One-Subject-Out(LOSO) cross-validation and report mean accuracy across subjects under two settings: complete multimodal input (EEG + eye movements) and missing modality (EEG only, with eye movements reconstructed by the generator).Results of comparison methods are reproduced from the respective original papers.}\label{tab1}
\end{table*}

\subsection{\textit{A Experiments Setup}}

\textbf{Dataset:} This experiment employs three emotion recognition datasets—SEED\cite{Lu2015}, SEED-IV,\cite{Zheng2019} and SEED-V\cite{seedv}—comprising EEG and eye movements bimodal signals from 12, 15, and 16 subjects, respectively, inducing 3, 4, and 5 emotional states. The EEG signals are extracted as 62 channels across 5 frequency bands, while the eye movements are represented as 33 (SEED, SEED-V) or 31 (SEED-IV) features, synchronously aligned with the EEG segments.

\textbf{Training Procedure and Hyperparameter Settings:} The model is trained in three alternating stages, with learning rates set to 1e-3 (stages 1 and 2) and 1e-4 (stage 3). These three stages are executed sequentially within every epoch and iterated over the entire training process. The main model is trained with a batch size of 2000 for 300 epochs, while the self-supervised adaptation uses a batch size of 256, triggered every 10 epochs with 50 internal steps per trigger. The loss balancing coefficients are set to \(\alpha = 0.5\), \(\beta = 0.7\), \(\eta = 0.2\), and \(\gamma = 0.1\). The detailed parameter sensitivity analysis and model structure are provided in the  supplementary material. 

\paragraph{\textbf{Baselines:}} We benchmark GUARD against these methods: DANN \cite{Ganin2016Domain}, DCCA\cite{seedv}, JDA \cite{Li2020Domain}, and MWACN \cite{MWACN}, CFDA-CSF \cite{CFDA-CSF}, MMDA \cite{MMDA}, CMSLNet \cite{CMSLNet}, MHESA \cite{MHESA}, CSMM \cite{CSMM}, CMGNN \cite{CMGNN}, MACDB\cite{MACDB}, SVM, TCA\cite{MMD}, DGCNN\cite{DGCNN}, RGNN\cite{RGNN}, BiDANN\cite{BiDANN}, BiHDM\cite{Bihdm}, PGCN\cite{PGCN}, UDDA\cite{UDDA}.

\subsection{\textit{B. Emotion Classification Performance}}

\textbf{1) Multimodal Results. } Tab.\ref{tab1}(left) reports the accuracy rates of each method on the SEED, SEED-IV and SEED-V datasets. The experimental results show: GUARD demonstrated the best performance across all three datasets. In the SEED-V five-classification task, the average accuracy of GUARD was 95.30\%, which was 4 percentage points higher than the existing most effective method MMDA, fully demonstrating the effectiveness of our framework in addressing coupled challenges such as cross-individual domain shift and modal absence. On SEED-IV, GUARD achieved an accuracy of 92.26\%, which was 2 percentage points higher than the second-best method CMGNN (90.21\%); on the SEED dataset, GUARD reached 94.99\%, approaching the level of CSMM (94.96\%).

\textbf{2) Missing-modal Results. }Tab.\ref{tab1}(right) reports the cross-subject emotion recognition accuracy rates of each method under the condition where only EEG signals are used in the target domain. GUARD achieved the best results on both the SEED and SEED-IV datasets, with average accuracy rates of 91.06\% and 77.33\% respectively. On SEED, GUARD outperformed the second-best BGAGCN-MT (89.66\%) by 1.40 percentage points; on SEED-IV, GUARD outperformed the second-best BGAGCN-MT (75.78\%) by 1.55 percentage points. This result indicates that even when the target domain lacks eye movement signals, GUARD can still learn cross-modal knowledge through the multimodal information of the source domain.

\subsection{\textit{C. Ablation Study}}
\begin{table}[t]
\centering

\renewcommand{\arraystretch}{1.1}
\setlength{\tabcolsep}{3pt}
\footnotesize
\begin{tabular}{lcclcc}\toprule
Multimodal& SEED & SEED-V & Missing-modal& SEED & SEED-V \\
\midrule

GUARD& 92.47& 96.12& GUARD& 89.84& 77.7\\
EEG zero            & 86.88& 76.07& w/o Eye enc.& 84.57& 70.82\\
Eye zero& 80.8& 94.44& $\mathcal{L}_{gen}$-only & 85.81& 73.29\\
w/o$\mathcal{L}_{adv}$& 87.26& 88.68& w/o g-mse& 85.63& 75.91\\
w/o $\mathcal{L}_{rec}$   & 90.17& 93.35& w/o $\mathcal{L}_{self}$ & 84.94& 73.82\\
\bottomrule
\end{tabular}
\caption{Ablation study results on SEED and SEED-V.}
\label{ablation}
\end{table}

\begin{figure}[t]
\centering
\includegraphics[width=0.9\linewidth]{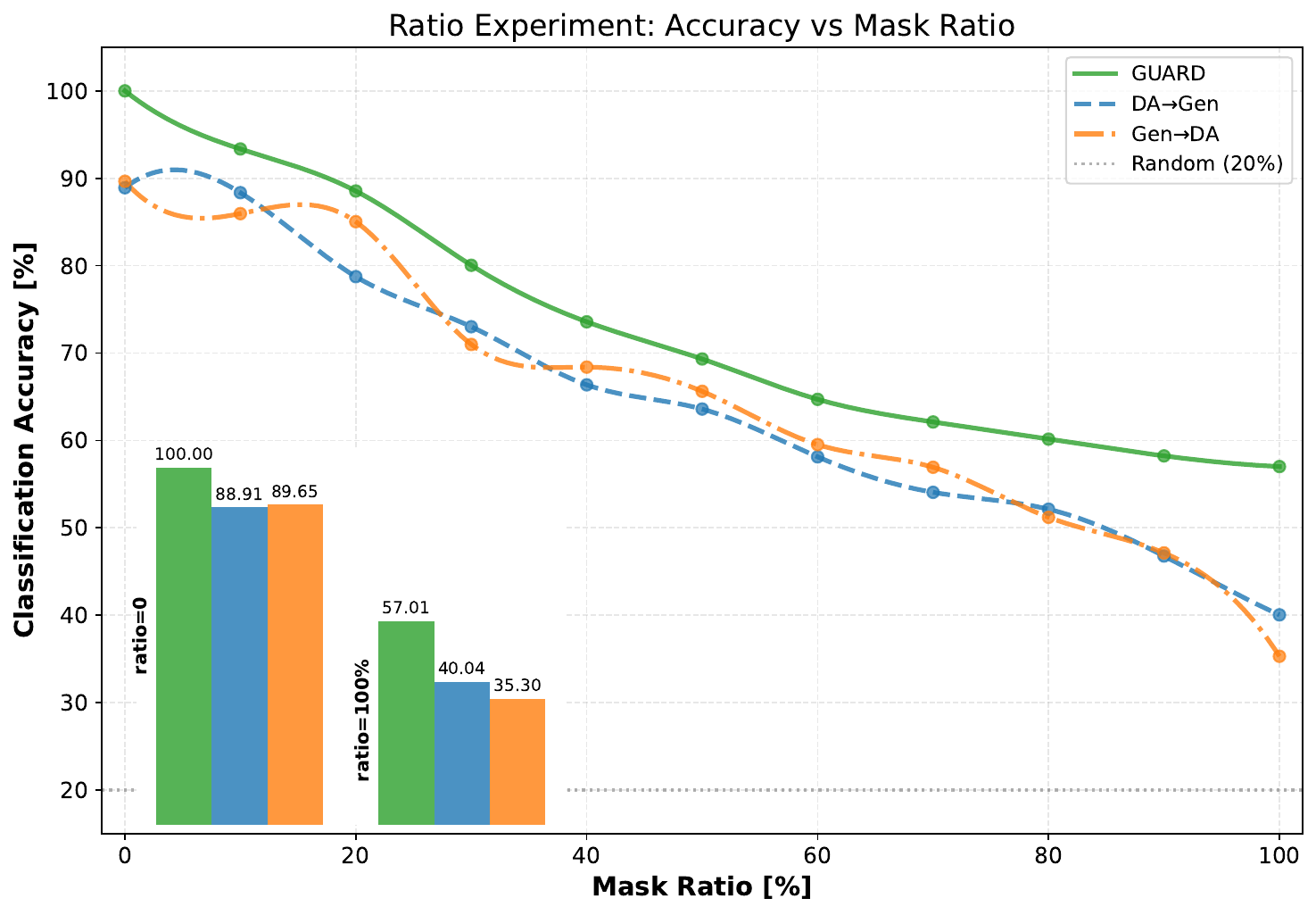} 
\caption{Accuracy comparison on the most challenging subject (SEED-V Session 2, Subject 6, exhibiting severe distribution shift) under varying target-domain eye-modality missing ratios. GUARD consistently outperforms Gen→DA and DA→Gen across all mask ratios. 
 }
\label{Ratio}
\end{figure}

\textbf{1) Remove component Results. }
Ablation experiments on SEED (session 2) and SEED-V (session 2) isolate each component's contribution (Tab.\ref{ablation}). Zeroing out eye signals (Eye zero) drops performance by 20.05\% on SEED-V, confirming the complementary role of eye movements. Completely removing the eye encoder (w/o Eye enc.) causes even worse results (84.57\% on SEED, 70.82\% on SEED-V), implying that discarding the stream is more detrimental than keeping a zero-input encoder that preserves fusion structure—supporting our hypothesis that cross-modal entanglement into the EEG anchor is vital for test-time missing modality. Training the generator only with signal fidelity (\(\mathcal{L}_{gen}\)-only) underperforms the full model, showing the need for perceptual supervision. Ablating adversarial alignment (w/o \(\mathcal{L}_{adv}\)) reduces performance, confirming its role in reducing subject-specific variation. Removing G-MSE (w/o g-mse) also causes drops, indicating uniform reconstruction weights cannot emphasize discriminative features. Lastly, omitting the self-supervised flow (w/o \(\mathcal{L}_{self}\)) significantly hurts performance, confirming that cycle-consistent calibration bridges the conditional distribution gap without target ocular labels.

\begin{figure}[t]
\centering
\includegraphics[width=0.9\linewidth]{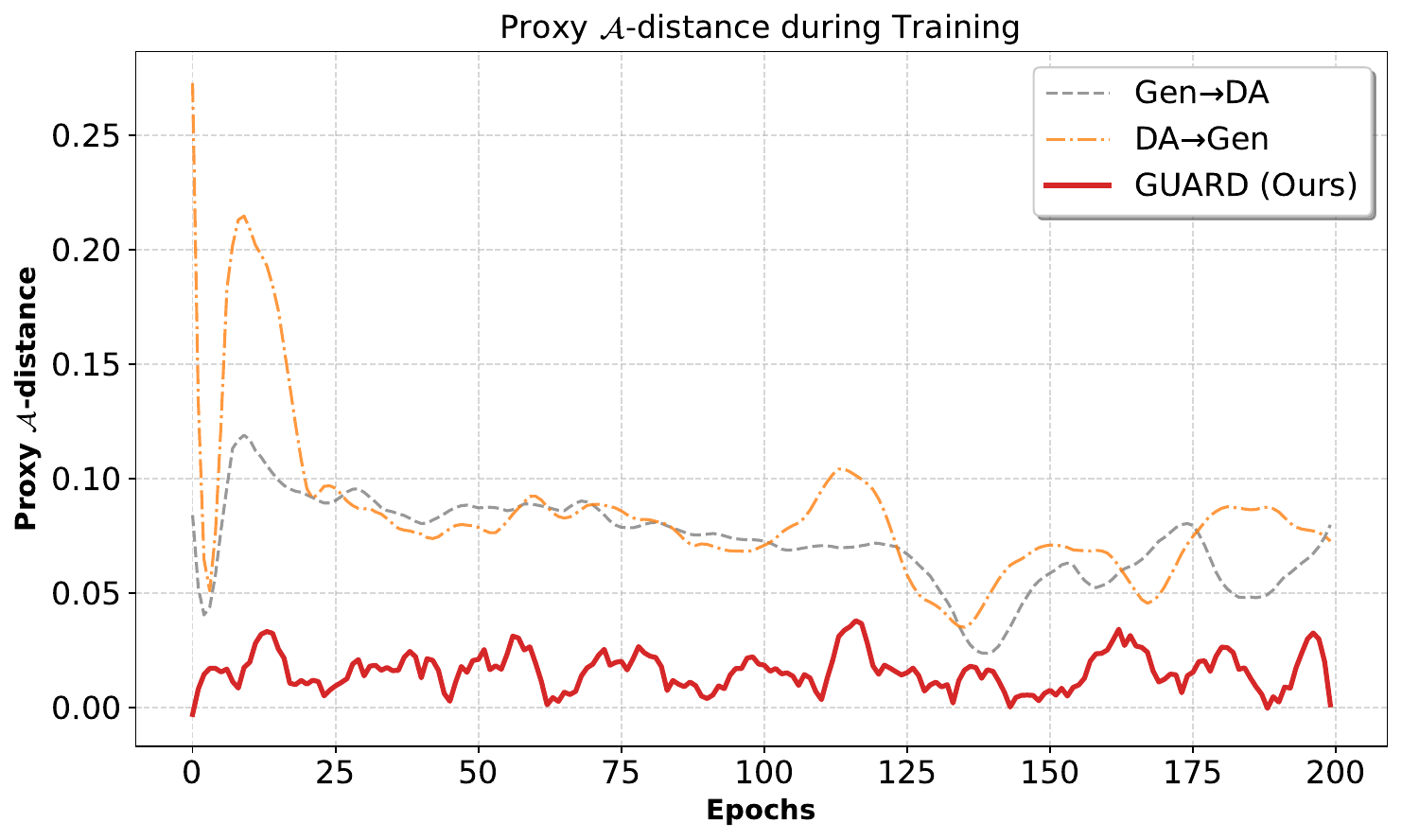} 
\caption{PAD curves of GUARD and two baseline methods (Gen→DA and DA→Gen) over training epochs. GUARD achieves the lowest and most stable domain discrepancy. 
 }
\label{PAD}
\end{figure}

\textbf{
2) Discard proportionally.} 
We evaluate GUARD under varying proportions of discarded eye movements in the target domain. As shown in Fig.\ref{Ratio}, GUARD outperforms Gen→DA and DA→Gen across all missing ratios, with the gap widening as modality loss increases—confirming that decoupled pipelines suffer from compounding error propagation, while our joint formulation effectively mitigates it through unified optimization. We further adopt Proxy A-distance (PAD)\cite{ben2010theory} to quantify domain discrepancy (Fig.\ref{PAD}). GUARD achieves the lowest and most stable PAD throughout training, while baselines exhibit significant divergence and instability. The periodic ripples in the GUARD curve coincide with episodic self-supervised triggers, indicating effective progressive domain alignment and robust convergence.

\subsection{\textit{D. Visualization}}

\begin{figure}[t]
\centering
\includegraphics[width=1\linewidth]{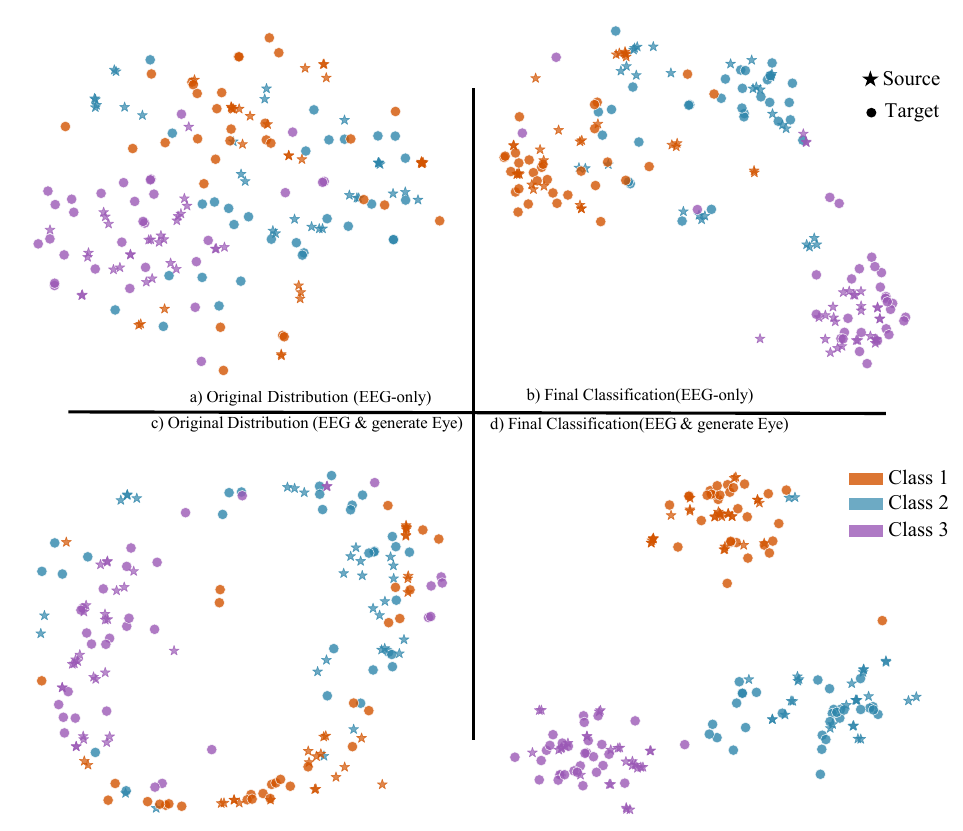} 
\caption{t-SNE-based feature distribution visualization for SEED Session 1 (Subject 7).}\label{scatter}

\end{figure}
\textbf{1) Distribution visualization. }In order to further explore the influence of the GUARD generation model on the distribution, we conducted a visualization experiment using t-SNE\cite{t-SNE} to verify its impact. As shown in Fig.\ref{scatter}, a clear domain shift exists between source and target when using EEG alone (a-b).  After incorporating the generated eye movements via GUARD (c-d), the domain discrepancy is substantially reduced, with source and target distributions becoming well-aligned. Notably, in the classification feature space (d), clusters are more compact and domain boundaries nearly vanish, demonstrating that GUARD effectively mitigates cross-domain distribution mismatch and enhances generalization through cross-modal generation and cyclic self-supervised training.

\begin{figure}[t]
\centering
\includegraphics[width=1\linewidth]{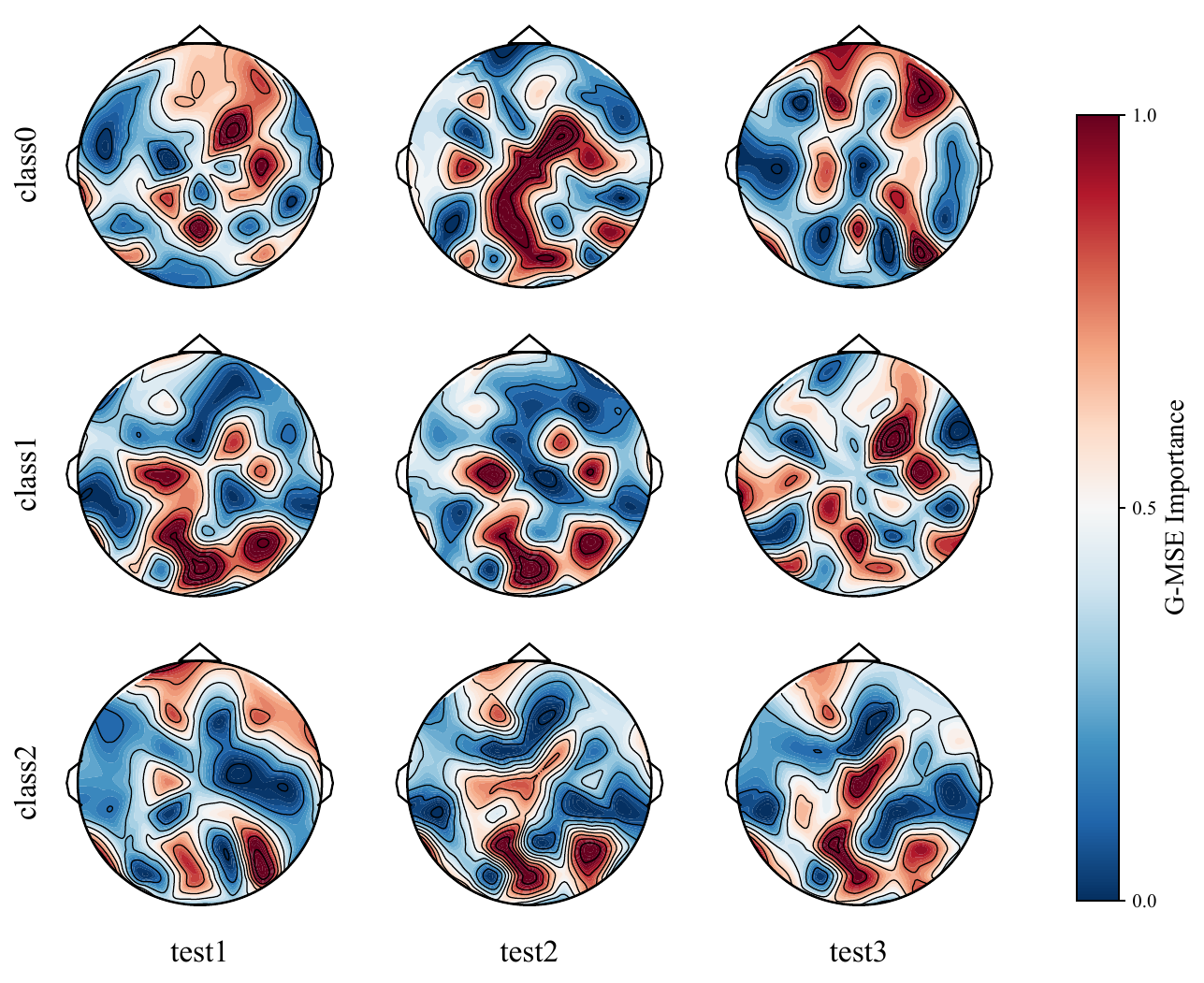} 
\caption{
Channel sensitivity analysis on SEED Session 1 (Alpha band). Three samples are randomly selected per emotion class. Heatmap values indicate gradient magnitude, with warmer colors reflecting higher decision sensitivity.  }
\label{fig5}
\end{figure}
\textbf{2) Importance analysis.} To verify that G-MSE can assign different weights to different samples, we conduct channel sensitivity analysis.  The experimental results are shown in Fig.\ref{fig5}. Notably, within the same class, sensitivity patterns vary considerably across samples, indicating that different samples rely on different channels even under the same label. Conversely, across different classes, similar sensitivity patterns can sometimes emerge. This suggests that channel importance is not solely label-determined but closely tied to sample-specific characteristics. Therefore, the generator must possess sample-adaptive channel awareness—dynamically focusing on the most informative channels for each sample rather than applying a uniform attention strategy. This observation directly motivates our sample-wise gradient weighting design for reconstruction guidance.

\section{Conclusion}

In this paper, we propose GUARD, a unified framework that reframes cross-subject adaptation and missing-modality completion from disjoint heuristics into a principled recovery and calibration of asymmetric joint distributions. By leveraging sample-adaptive gradient weighting ($\mathcal{L}_{\text{G-MSE}}$), GUARD establishes a task-discriminative EEG anchor manifold that preemptively entangles cross-modal semantics. To resolve the structural collapse caused by missing ocular signals, GUARD explicitly recovers the conditional distribution $P(x_{\text{eye}} \mid x_{\text{eeg}})$ via perceptual constraints and reformulates target-domain adaptation as an overdetermined inverse problem. Through a cycle-consistent self-supervised flow with Lagrangian latent regularization, the recovered joint space is effectively calibrated directly on the target manifold without requiring target ocular labels. Comprehensive evaluations on the SEED series datasets demonstrate that GUARD eliminates cascading error propagation, offering a resilient foundation for real-world affective computing under unpredictable sensor failures.

\bibliography{aaai2027}


\end{document}